\documentclass[twocolumn,amsmath,amssymb]{revtex4}
\usepackage{graphicx}
\usepackage{hyperref}
\usepackage{natbib}

\renewcommand{\epsilon}{\varepsilon}
\renewcommand{\phi}{\varphi}

\begin{document}
\begin{titlepage}
	\title{Gravitational distortion on photon state at the vicinity of the Earth}
	\author{Qasem Exirifard}
	\email{qexirifa@uottawa.ca}

	\author{Ebrahim Karimi}
	\email{ekarimi@uottawa.ca}
	
	\affiliation{Department of Physics, University of Ottawa, 25 Templeton St., Ottawa, Ontario, K1N 6N5 Canada}
\
	\begin{abstract}
As a photon propagates along a null geodesic, the space-time curvature around the geodesic distorts its wave function. We utilise the Fermi coordinates adapted to a general null geodesic, and derive the equation for interaction between the Riemann tensor and the photon wave function. The equation is solved by being mapped to a time-dependent Schr\"odinger equation in $(2+1)$ dimensions. The results show that as a Gaussian time-bin wavepacket with a narrow bandwidth travels over a null geodesic, it gains an extra phase that is a function of the Riemann tensor evaluated and integrated over the propagation trajectory. This extra phase is calculated for communication between satellites around the Earth, and is shown to be measurable by current technology.
	\end{abstract}
	
	\maketitle
	
\end{titlepage}

\section{Introduction} 
Einstein gravity is a geometric theory for gravity wherein energy/mass distribution curves its surrounding space-time geometry and particles 
propagate along the geodesics of the curved geometry. The light bending by a gravitational source, manifesting that photons propagate along null geodesics, was first observed by Eddington et al.~\cite{Edington} with $30\%$ precision, and the effect has now been measured with the precision of $0.01\%$~\cite{Lebach:1995zz,Shapiro:2004zz,Fomalont_2009,Lambert_2009}.  In light bending as well as all other observed effects of Einstein gravity, the photon is treated as a point-like particle\cite{GRtest}.   The photon, however, is governed by the rules of quantum mechanics  where particle-wave duality is manifest. As a photon moves along the geodesic, its quantum wave function interacts with the curvature of the space-time geometry around the geodesic and gets distorted. Whenever the photon wave function is used as an information carrier~\cite{Ursin:07,Yin:17,Yin:17PRL,2020npjQI,2020NJPh22i3074H}, the distortion affects the communication channel and may introduce errors. The current race to establish a quantum network in space~\cite{QuantumCommunication,Dequal_2021,yin2017satellitebased} may need to take into account how the curvature of the space-time geometry affects the wave function of the photon as it moves along a geodesic. Approximations, where all the multi-polar modes are neglected, are used to calculate the Green function for the propagation, which resulted in a measurable distortion~\cite{Bruschi:2013sua, Bruschi:2014cma, Bruschi:2021all,Bruschi:2021rhk}. Jonsson et al.~\cite{Jonsson:2020npo} has kept the first few multi-polar modes to calculate the distortion for a photon scattered from a black hole. Though none of these methods return the exact distortion, they highlight that the distortion is a substantial effect. In Ref.~\cite{Exirifard:2020yuu}, the Fermi coordinates along a geodesic is adapted, and a distortion of measurable magnitude is reported for communication between the Earth and the International Space Station. However, turbulence due to the Earth's atmosphere adds noise, and may not allow measurement of the effect. Here, we aim to calculate the gravitational distortion for communication performed between two satellites around the Earth where atmospheric effects are absent. The paper is organised as follows: Section~\ref{Section2} considers a photon that propagates along a null geodesic in a general curved space-time geometry. The photon is approximated to a time-bin wave packet with a very narrow frequency line. The Fermi-coordinates adapted to a null geodesic are utilised to calculate the interaction between the photon's wave function and the curvature of the space-time geometry around the geodesic. We show that as the photon propagates along the null geodesic, it gains an extra phase that is a function of the Riemann tensor evaluated and integrated over the null geodesic. This extra phase, which depends on the space-time geometry, is calculated for communication between satellites near the Earth. The results are shown in Section~\ref{Section3}, and Section~\ref{Section4} provides the discussion on how to measure this extra phase.

\section{Predicting a general geometric phase}
\label{Section2}
The photon moves on the null geodesic $\gamma$. We adapt the Fermi frame wherein the metric on the geodesic coincides to the Minkowski metric, and Levi-Cevita symbol vanishes too:
\begin{eqnarray}
g_{\mu\nu}|_\gamma = \eta_{\mu\nu},\\
\Gamma^{\mu}_{\nu\eta} |_\gamma= 0.
\end{eqnarray}
We represent the coordinates in the Fermi frame by $(x^{\pm},x^1,x^2)$ where  $x^{\pm}=\frac{1}{\sqrt{2}}(x^3\pm ct)$
are the Dirac light-cone coordinates while $x^+$ is tangent to $\gamma$~\cite{Dirac:49}. 
The metric around the geodesic in the Fermi coordinates up to quadratic order in the transverse coordinates  is given by~\cite{Blau:06}:
\begin{eqnarray}
    \label{Eq1}
	ds^2 &=& 2 dx^+ dx^- + \delta_{ab}\, dx^a dx^b-R_{+\bar{a}+\bar{b}} x^{\bar{a}} x^{\bar{b}} (dx^+)^2\nonumber\\
	&-&     \frac{4}{3}R_{+\bar{b}\bar{a}\bar{c}} x^{\bar{b}}x^{\bar{c}} (dx^+ dx^{\bar{a}})-\frac{1}{3} R_{\bar{a}\bar{c}\bar{b}\bar{d}} x^{\bar{b}}x^{\bar{c}} (dx^{\bar{a}} dx^{\bar{b}})\nonumber\\
	&+& O(x^{\bar{a}} x^{\bar{b}} x^{\bar{c}})\,,
\end{eqnarray}
where $x^a=(x^1,x^2)$, and $x^{\bar{a}}=(x^{-},x^a)$,  and all the curvature components ($R_{+\bar{a}+\bar{b}}$, $R_{+\bar{a}\bar{c}\bar{d}}$ and $R_{\bar{a}\bar{b}\bar{c}\bar{d}}$) are evaluated on $\gamma$, and Einstein's notation is used wherein twice appearance of an index variable in a single term means summation over that index, and $\delta_{ab}$ is the Kronecker delta.  We approximate the space-time geometry around the Earth to the Schwarzschild space-time geometry. Therefore, the effective Lagrangian of a massless point particle propagating on a null geodesic can be given by,
\begin{equation}
	{\cal L} = -\left(1-\frac{m}{r}\right)\,\dot{t}^2+ \frac{\dot{r}^2}{1-\frac{m}{r}} + r^2 \left(\dot{\theta}^2+ \sin^2\!\theta\, \dot{\phi}^2\right)\,,
\end{equation}
where dot presents variation with respect to the affine parameter on the geodesic, $m={\left(2 G_N M_{\oplus}\right)}/{c^2}$, and $M_{\oplus}$ is the mass of the Earth. We choose the units such that the speed of light in vacuum is set to 1, i.e. $c=1$, and $m=1$. Due to the spherical symmetry, without loss of generality, we can choose the equatorial plane $\theta={\pi}/{2}$ and $\dot{\theta}=0$ to describe any given geodesic at all time. The cyclic variables of $\phi$ and $t$ lead to invariant quantities: $r^2 \dot{\phi} = l$, $(1-{1}/{r}) \dot{t} = E$, where $l$ and $E$ are constant values. We consider null geodesics reaching the asymptotic infinity, and set $E=1$. Due to the form of the Lagrangian, its Legendre transformation, which is the Lagrangian itself, is invariant. We consider a null geodesic, and set ${\cal L}=0$ that gives $|\dot{r}|=\sqrt{1 - \frac{1}{r^2}\left(1-\frac{1}{r}\right) l^2}$.  Let $\hat{e}_t$, $\hat{e}_r$, $\hat{e}_\phi$ and $\hat{e}_\theta$ represent the normalized unit vectors in $t,r,\theta,\phi$ coordinates. The normalized unit vectors in the Fermi coordinates can be chosen as
\begin{subequations}
\begin{eqnarray}
\label{5a}
	\hat{e}_+&=&  \frac{f}{\sqrt{2}}(+\hat{e}_t+   \dot{r}\hat{e}_r + rf \dot{\phi}\hat{e}_\phi),\\
\label{5b}	
	\hat{e}_- &=&\frac{1}{\sqrt{2}f}(-\hat{e}_t+  \dot{r} \hat{e}_r + r f \dot{\phi} \hat{e}_\phi)\,,\\
\label{5c}	
	\hat{e}_1 &=& - r f \dot{\phi} \hat{e}_r +\dot{r} \hat{e}_\phi \,,\quad
	~\hat{e}_2 = \hat{e}_\theta\,,
\end{eqnarray}
\end{subequations}
where $f=\sqrt{1-1/r}$. The components of the Riemann tensor in the Fermi coordinates should be computed by the tensor transformation law \cite{Manasse:1963zz}:
\begin{equation}
    R_{\alpha\beta\gamma\delta}= R_{\mu'\nu'\sigma'\tau'}
    (\hat{e}_\alpha)^{\mu'} 
    (\hat{e}_\beta)^{\nu'}
    (\hat{e}_\gamma)^{\sigma'}
    (\hat{e}_\delta)^{\tau'}.
\end{equation}
Utilizing the abstract method employed in \cite{Exirifard:2020yuu} identifies the non-vanishing components of the Riemann tensor in the Fermi frame 
\begin{subequations}\label{Eq5}
	\begin{eqnarray}
		R_{+-+-}&=& \frac{3 l^2(r-1)}{2r^6}-\frac{1}{r^3}\,,\label{Rpmpm} \\
		R_{+2+2} &=&-R_{+1+1}=\frac{3l^2}{4r^5}\,,\\
		R_{+-+1} &=& \frac{3\sqrt{2} l}{2r^4}   \dot{r}\,.
	\end{eqnarray}
\end{subequations}
Next, we consider the electromagnetic potential $A_\mu$, whose field strength is given by $F_{\mu\nu}= \partial_\mu A_\nu - \partial_\nu A_\mu$. The dynamics of the electromagnetic potential in a curved space-time geometry endowed with metric $g_{\mu\nu}$ around the geodesic, and is given by,
\begin{subequations}
\label{Eq7}
\begin{eqnarray}
	\Gamma[{A}_\mu] &=& -\frac{1}{4} \int\!\!d^4\!x\,g^{\mu\mu'\nu\nu'} F_{\mu\nu} F_{\mu'\nu'}\,,\\
	g^{\mu\mu'\nu\nu'}& =&\frac{1}{2}\sqrt{-\det g}\, (g^{\mu\mu'}g^{\nu\nu'}-g^{\mu\nu'}g^{\nu\mu'}) \,.
\end{eqnarray}
\end{subequations}
Here, $g^{\mu\nu}$ is the inverse of the metric, and $\det g$ is its determinant. Note that $g^{\mu\mu'\nu\nu'}$ has all the symmetries of $F_{\mu\nu} F_{\mu'\nu'}$ under exchange of its indices. The functional variation of the action with respect to the gauge field gives its equation of motion, i.e.,
\begin{equation}
	\label{Eq8}
	\partial_\mu \left(g^{\mu\mu'\nu\nu'} F_{\mu'\nu'}\right)=0\,,
\end{equation}
which we would like to solve for a photon that travels along a null geodesic. We choose the Fermi coordinates adapted to the null geodesic, Eq.~\eqref{Eq1}, to describe the space-time at the vicinity of the geodesic where the components of the Riemann tensor are given in Eq.~\eqref{Eq5}. At the vicinity of the Earth, the components of the Riemann tensor are minimal. We, therefore, treat them as a perturbation. We introduce $\epsilon$ as the systematic parameter of the perturbation. In other words, we add a factor of $\epsilon$ to all terms in Eq. \eqref{Eq1} where the components of the Riemann tensor are present. At the end of the computation, we set $\epsilon=1$. The $\epsilon$-perturbation to the electromagnetic potential and $g^{\mu\mu'\nu\nu'}$ follow:
\begin{subequations}
\label{Eq9}
\begin{eqnarray}
	A_{\mu} &=& A^{(0)}_\mu + \epsilon A^{(1)}_\mu   + O(\epsilon^2) \,,\\
	g^{\mu\mu'\nu\nu'} &=& g^{(0) \mu\mu'\nu\nu'}+ \epsilon g^{(1)\mu\mu'\nu\nu'} + O(\epsilon^2) \,.
\end{eqnarray}
\end{subequations}
Equation~\eqref{Eq8} at the leading order in $\epsilon$ can be simplified to
\begin{eqnarray}
	\label{Eq10}
	\Box^{(0)}  A^{(0)\mu} + \partial^{\mu} \partial_\nu A^{(0)\nu} = 0.
\end{eqnarray}
Henceforth $\eta^{\mu\nu}$ is utilized to move up or down the indices, i.e., $A^{(0)\nu}=\eta^{\nu\lambda}A^{(0)}_{\lambda},~\partial^\mu = \eta^{\mu\nu} \partial_\nu,$ where $\eta^{\mu\nu}$ represents the Minkowski metric in Dirac coordinates. We choose the Lorentz gauge, 
$\partial_\nu A^{(0)\nu} = 0$, which simplifies the equation for $A^{(0)}$ to $\Box^{(0)} A^{(0)}_\mu = \left(2 \partial_+ \partial_- + \nabla^2_{\perp}\right) A^{(0)}_\mu=0$ where $\nabla_\perp^2$ is the Laplace operator in $x^1$ and $x^2$ directions: $\nabla_\perp^2 = \partial_1^2 + \partial_2^2$. Utilizing the Fourier expansion of the gauge field in terms of the variable $x^-$, i.e., $A^{(0)}_\mu =\int\! d\omega  f^{(0)}_\mu(\omega, x^+,x^a) e^{i\omega x^-}$ leads to,
\begin{equation}
	\label{Eq12}
	(2 i \omega \partial_+ + \nabla_\perp^2) f^{(0)}_\mu= 0,
\end{equation}
which is referred to as the paraxial Helmholtz equation. Note that due to the definition of $\hat{e}_-$ in Eq.~\eqref{5b}, the gravitational red-shift is already encoded in the Fourier expansion of $A^{(0)}_\mu$. We refer to $f^{(0)}_\mu(w, x^+,x^a)$ as the structure function of the photon with frequency $\omega$ in mode $\mu$. The structure mode can be expanded in terms of the Hermite-Gaussian or Laguerre-Gaussian modes. For the purpose of communication, we are interested in a field configuration that can be understood as a perturbation modulated over a frequency that holds: 
\begin{equation}
	\label{Eq13}
	|\partial_+ f^{(0)}_\nu|\ll \omega |f^{(0)}_\nu|,\qquad 	|\partial_- f^{(0)}_\nu|\ll \omega |f^{(0)}_\nu|
\end{equation}
which is the same as the paraxial approximation in optics. Employing Eq.~\eqref{Eq13} in the Lorenz gauge condition yields:
\begin{equation}
	\label{Eq14}
	\omega f^{(0)}_+ + \partial_+ f_-^{(0)}+  \partial_1  f_1^{(0)}+  \partial_2  f_2^{(0)}=0.
\end{equation}
The paraxial approximation conditions, i.e. Eq.~\eqref{Eq13}, imply the following perturbative solutions:
\begin{subequations}
	\label{Eq15}	
	\begin{eqnarray}
		f^{(0)}_+&=&0\,,\\
		\label{Eq15b}
		\partial_+ f_-^{(0)} +\partial^a f^{(0)}_a&=&0\,,
	\end{eqnarray}
\end{subequations}
where $\partial^a f^{(0)}_a= \partial_1 f^{(0)}_1+\partial_2 f^{(0)}_2$
is used. We solve Eq.~\eqref{Eq15b} for $ f_-^{(0)}$. This leaves $f_a^{(0)}$ as the physical modes, which can be perceived as the distribution of the photon's polarization. This means that each polarization of photon that we choose to represent by $\Psi$, satisfies Eq.~\eqref{Eq12}. The paraxial wave equation, Eq.~\eqref{Eq12}, can be rewritten as $-\frac{1}{2\omega} \nabla_\perp^2 \Psi= i\partial_+ \Psi$, which is the Schr\"odinger equation for a particle with a rest mass of $\omega$ in``2+1" dimensions where $x^+$ plays the role of time.

\begin{figure}[t]
	\begin{center}
		\includegraphics[width=0.45 \textwidth]{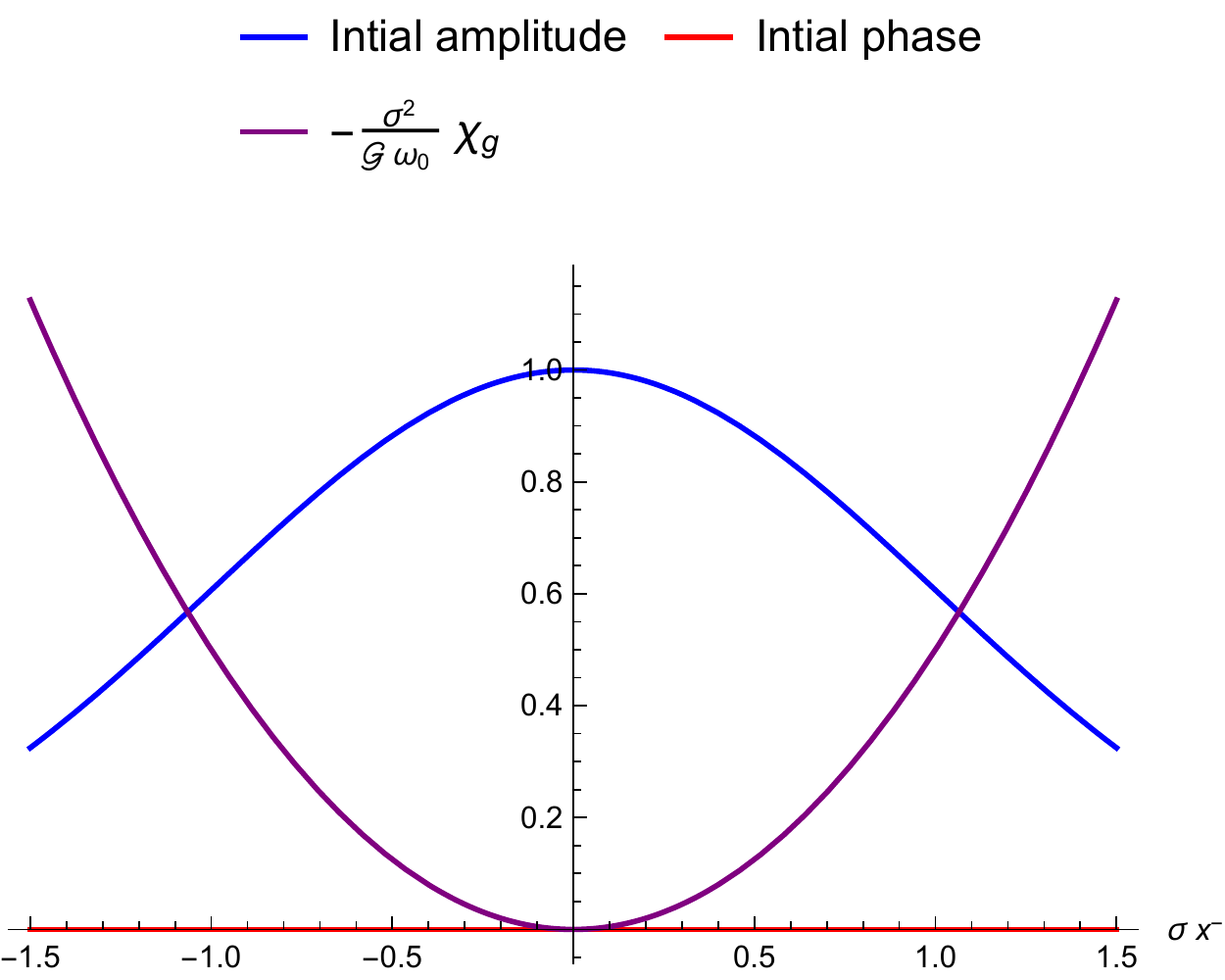}
	\end{center}
\caption{The initial amplitude (shown in blue), the initial phase which chosen to be zero (shown in red), and the geometric phase multiplied by $-\frac{\sigma^2}{\omega_0 {\cal G}}$ (shown in purple) in term of $\sigma x^-$.}
	\label{Fig0}	
\end{figure}

Utilizing Eqs.~\eqref{Eq9} and \eqref{Eq10} in Eq.~\eqref{Eq8} yields the equation of motion for $A^{(1)}$,
\begin{equation}
	\label{Eq17}
	\Box^{(0)} A^{(1)\mu} = -\partial_\nu\left( g^{(1)\mu\mu'\nu\nu'} F^{(0)}_{\mu'\nu'}\right)\,,
\end{equation}
where Lorenz gauge condition is assumed on $A^{(1)}$ too. The propagation of a photon in a smooth space-time geometry holds $|\partial_\lambda g^{(1)\mu\nu}|\ll \omega |g^{(1)\mu\nu}|$. Therefore,  the derivative of the  components of the metric on the right hand side of Eq.~\eqref{Eq17} can be neglected, and thus we have $\Box^{(0)} A^{(1)\mu}= -g^{(1)\mu\mu'\nu\nu'} \partial_\nu F^{(0)}_{\mu'\nu'}$. We express $F^{(0)}_{\mu'\nu'}$ in terms of $A^{(0)}_{\mu'}$:
\begin{equation}
	\label{Eq20}
	\Box^{(0)} A^{(1)\mu}   = -g^{(1)\mu\mu'\nu\nu'}   (\partial_\nu \partial_{\mu'} A_{\nu'} - \partial_\nu \partial_{\nu'} A_{\mu'})\,,
\end{equation}
The paraxial approximation expressed in Eq.~\eqref{Eq13} implies that the dominant term on the right hand side of Eq.~\eqref{Eq20} is the one that $\partial_-^2$ acts on $A^{(0)}$. Keeping only the dominant term results 
\begin{equation}
	\label{Eq21}
	\Box^{(0)} A^{(1)\mu}   = \left(g^{(1)--\mu\alpha}-g^{(1)-\alpha\mu-}\right)  \partial_-^2 A_\alpha^{(0)}\,.
\end{equation}
Due to gauge symmetry and the chosen Lorentz gauge, we choose to solve Eq.~\eqref{Eq21} for $\mu=-, 1,2$. Substituting Eq.~\eqref{Eq15} into Eq.~\eqref{Eq21} results in 
\begin{subequations}
	\label{Eq22}
\begin{eqnarray}
    \label{Eq22a}
	\Box^{(0)} A^{(1)}_+ &=& 0,\\  
	\label{Eq22b}
	\Box^{(0)} A^{(1)}_a &=&  - g^{(1)--} \partial_-^2 A_{a}^{(0)}.
\end{eqnarray}
\end{subequations}
$g^{(1)--}$ can be expressed in terms of the components of the Riemann tensor, and thus we have,
\begin{eqnarray}
	\label{Eq23}
	\Box^{(0)} A^{(1)}_i  =   - R_{+\bar{a}+\bar{b}}~x^{\bar{a}} x^{\bar{b}}\partial_-^2 A^{(0)}_i .
\end{eqnarray}
We would like to consider the Fourier transformation of $A^{(1)}_i$ with respect to the variable $x^-$, i.e., $A^{(1)}_i =\int d\omega  f^{(1)}_i(\omega, x^+,x^a) e^{i\omega x^-}$, where $f^{(1)}_i$ is the correction to the structure function of mode $i$. Utilizing the Fourier transformations in Eq.~\eqref{Eq23} yields,
\begin{eqnarray}
	\label{Eq24}
	&&(2i \omega \partial_+ + \nabla_\perp^2) f^{(1)}_i =\nonumber\\  &&\left(- R_{_{+-+-}} \partial_\omega^2  +2 i   R_{_{+-+-a}}  x^a \partial_\omega+ \omega^2 R_{_{+a+b}} x^a x^b\right)\omega^2 f_i^{(0)}.\nonumber\\
\end{eqnarray}
We observe that different physical modes, i.e., different $i$, are not coupled at the sub-leading order. Therefore, without loosing generality, we consider one physical mode, and we set $i=1$. However, all the results that we will calculate, will be equally valid for $i=2$. We consider $f_1^{(0)}$ in the form of $f_1^{(0)}= A(\omega) f_{mn}(\omega, x^+, x^a)$, where $f_{mn}$ is the Hermite-Gaussian mode, and $A(\omega)$ is the amplitude that we choose as a normal distribution around $\omega=\omega_0$ with the width of $\sigma$ for the first polarization of photon: $A(\omega)= \frac{1}{\sqrt{\sigma}\pi^{\frac{1}{4}}} \exp\left(-\frac{(\omega-\omega_0)^2}{2 \sigma^2}\right)$. 
We consider a time-bin wavepacket with a narrow bandwidth such that the first term on the right hand side of Eq.~\eqref{Eq24} is the dominant term. Since $\sigma$ is very small, we can utilise $f_1^{(0)}\approx A(\omega) f_{mn}(\omega_0, x^+, x^a)$. This approximation allows us to neglect $\partial_\omega f_{mn}$ in derivatives of $f_1^{(0)}$ with respect to $\omega$. Size of the wave packet is identified by two parameters. Its size perpendicular to its trajectory is given by the width of the beam while its size in direction of propagation is proportional to $\frac{c}{\sigma}$. We assume $\frac{c}{\sigma}$ is much larger than the width of package, and the wavepacket is extended in the direction of propagation where the first term on the right hand side of Eq.~\eqref{Eq24}, becomes the dominant term. Keeping only the dominant term yields 
\begin{eqnarray}
	\label{Eq27}
	&&(2i \omega \partial_+ + \nabla_\perp^2) f^{(1)}_1 =\nonumber\\
	&&- R_{+-+-} \left(\frac{\omega ^2 (\omega -\omega_0)^2}{\sigma^4}+\frac{\omega  (4 \omega_0-5 \omega )}{\sigma ^2}\right)\! f_1^{(0)}\!,
\end{eqnarray}
where $2 f_1^{(0)}$ on the right hand side of Eq.~\eqref{Eq27} is also neglected. Equation~\eqref{Eq27} can be perceived as the perturbation of
\begin{eqnarray}
	\label{Eq28}
	\left(2i \omega \partial_+ + \nabla_\perp^2\right) \Psi= 2 \epsilon \omega V(x^+) \Psi,
\end{eqnarray}
where 
\begin{subequations}
\label{Eq29}
\begin{eqnarray}
	\label{Eq29a}
	\Psi&=& f_1^{(0)}+ \epsilon f_1^{(1)}+O(\epsilon^2),\\
	\label{Eq29b}
	V(x^+)&=& -  \left(\frac{\omega (\omega -\omega_0)^2}{2\sigma^4}+\frac{4 \omega_0-5 \omega }{2\sigma ^2}\right) R_{+-+-}(x^+).\nonumber\\
\end{eqnarray}
\end{subequations}
Equation~\eqref{Eq28} can be rewritten as,
\begin{equation}
	\label{Eq30}
	i\partial_+ \Psi = \left(-\frac{1}{2\omega} \nabla^2 +\epsilon V(x^+)\right) \Psi,
\end{equation}
which is the Schr\"odinger equation for a particle with a mass ``$\omega$'' in $(2+1)$ dimensions with a time-dependent potential where the potential is only a function of time. For any $\Psi^{(0)}$ that satisfies $i\partial_+ \Psi^{(0)} = -\frac{1}{2\omega} \nabla^2  \Psi^{(0)}$, the perturbative solution to Eq.~\eqref{Eq30} is $\Psi=\Psi^{(0)} \left(1+ \epsilon \chi(x^+)\right)$ where $\chi = -i \int^{x^+}_0 d\tau\, V(\tau)$. This implies that the correction to the structure function, $f^{(1)}_1$, is expressed in term of the structure function, $f^{(0)}_1$, i.e.,
\begin{eqnarray}
	\label{Eq31}
	f^{(1)}_1 =  i f^{(0)}  \left(\frac{\omega (\omega -\omega_0)^2}{2\sigma^4}+\frac{4 \omega_0-5 \omega }{2\sigma ^2}\right) {\cal G},
\end{eqnarray}
where $\cal G$, the geometrical factor, is given by 
\begin{eqnarray}
	\label{Eq32}
	{\cal G}=  \int_0^{x^+} d\tau R_{+-+-}(\tau).
\end{eqnarray}
Here, $\tau$ is the affine parameter on the geodesic and $x^+=0$ is the wave-packet initial plane. Recalling that $f^{(1)}_1$ and $f^{(0)}_0$ are the Fourier transformation of $A^{(1)}_1$ and $A^{(1)}_0$, allows us to integrate over $\omega$, and obtain the electromagnetic field:
\begin{eqnarray}
	\label{Eq33}
	A_1  &=& \sqrt{2\sigma} \pi^{\frac{1}{4}}  e^{-\frac{(\sigma x^-)^2}{2}+ i \omega_0 x^-} f_{mn}(\omega_0, x^+, x^a)\nonumber\\
	&\times&\left(1- i \frac{\omega_0 {\cal G}}{2 } ( x^-)^2\right),  
\end{eqnarray}  
where only the dominant term is kept, and we set $\epsilon=1$ (note that $\epsilon$ is a dummy parameter to systematically track the perturbation). We could have chosen $i=2$ to obtain the same expression for the second polarization, i.e., $A_2$. We, therefore, observe that as a Gaussian time-bin wave-packet, with sharp width of $\sigma$ around frequency of $\omega_0$, travels over the geodesic, and it gains an extra geometric phase that is given by, 
\begin{equation}
	\label{Eq34.01}
	\chi_{g}=  -  \frac{\omega_0 {\cal G}}{2 } ( x^-)^2,
\end{equation}
where $\cal G$ is the integration of ``${+-+-}$'' component of the Riemann tensor evaluated on the geodesic, as defined in Eq.~\eqref{Eq32}. 
Equation~\eqref{Eq34.01} is in accord with~\cite{Exirifard:2020yuu} wherein the equations are solved by a different method. Let it be emphasized that $\chi_g$ is the change in the phase of a Gaussian beam with the width of $\sigma$. There exist some difficulties associated with measuring $\chi_g$ at the far tail ($|\sigma x^-|\geq 5$) of the Gaussian beam because the amplitude decreases exponentially, and it would not be easy to generate a  Gaussian beam whose far tail remains Gaussian too. To avoid these problems, we suggest to measure $\chi_g$ around the peak of the Gaussian beam, or equivalently for $ |\sigma x^-| \lesssim 1$. In doing so, it is convenient to re-express $\chi_g$ to 
\begin{equation}
	\label{Eq34}
	\chi_{g}=  -  \frac{\omega_0 {\cal G}}{2 \sigma^2} (\sigma x^-)^2,
\end{equation}
and note that $\chi_g$ is measured for $|\sigma x^-|\lesssim 1$. Equation~\eqref{Eq34} explicitly shows that the maximum measurable value of $\chi_g$ depends on $\sigma$. Figure~\ref{Fig0} depicts the amplitude, the initial phase and the change in the phase in term of $\sigma x^-$ for $-1.5\leq \sigma x^-\leq 1.5$.

%
%
\begin{figure*}
  \centering
  \begin{tabular}{c @{\hspace{1.5cm}} c }
    \includegraphics[width=.44\linewidth]{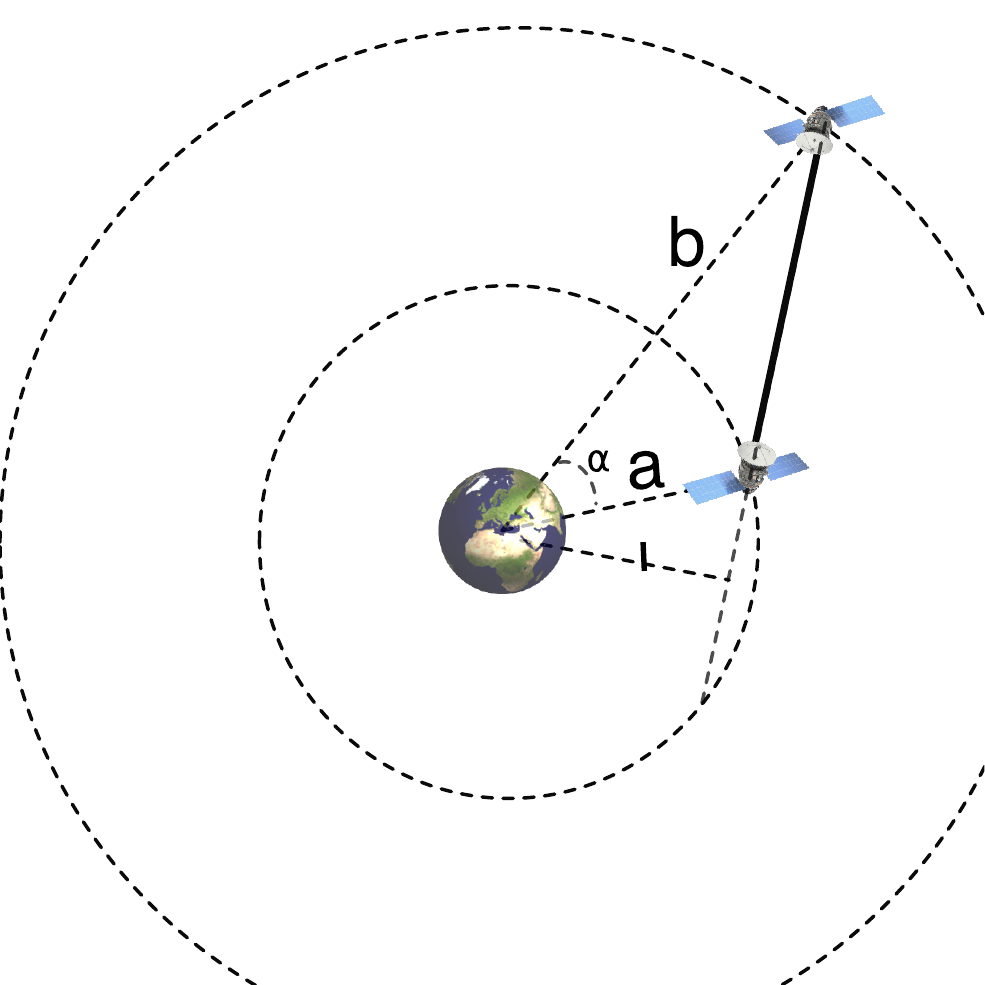} &
    \includegraphics[width=.44\linewidth]{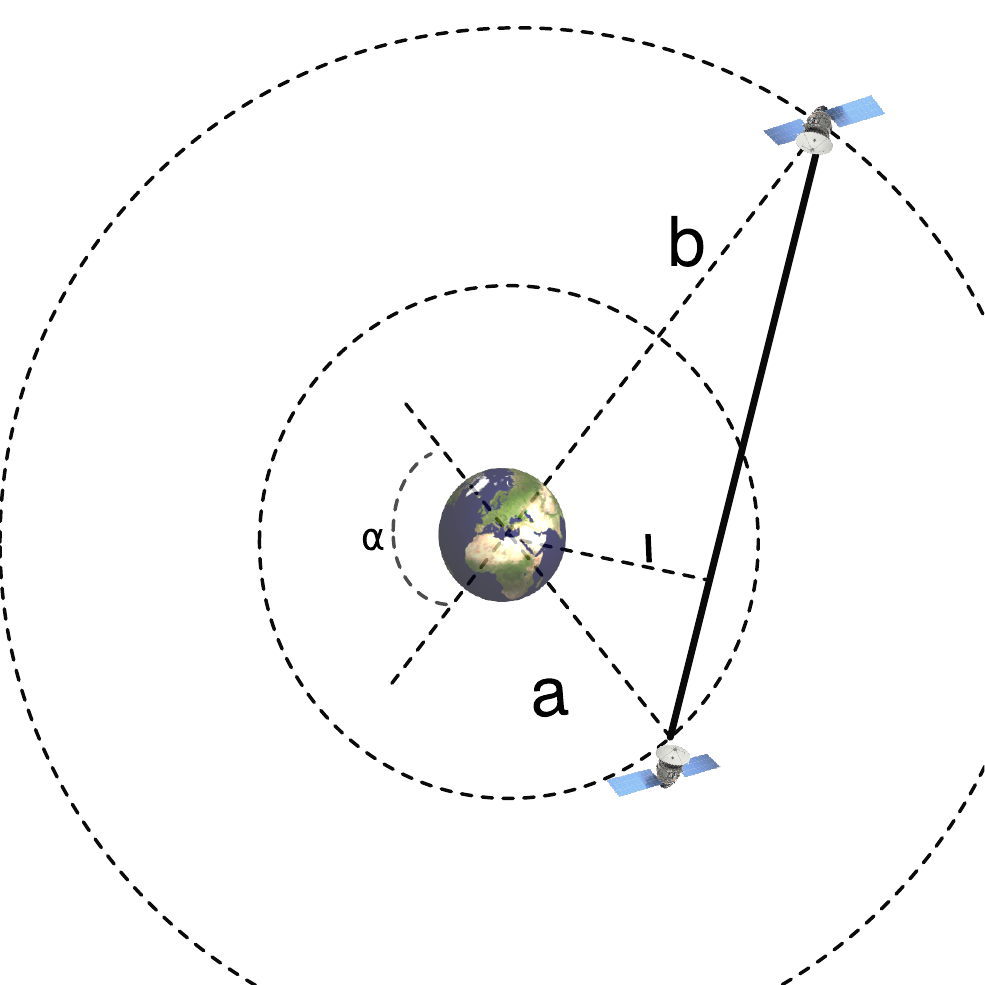} \\
    \small (a) During the entire journey of the signal  $\dot{r}>0$. & \small (b) At the beginning $\dot{r}<0$ then $\dot{r}>0$.
  \end{tabular}
  \caption{\textbf{Alice at radius $a$, at one instant, sends a signal toward Bob at radius $b$ with $a<b$}. The signal propagates along a null geodesic that can be approximated to a straight line. Notice that $l$ is the minimum distance between the centre of the Earth and the line, or the extrapolation of the line, connecting Alice and Bob. $\alpha$ is the angle between lines connecting Alice and Bob to the centre of the Earth. Alice and Bob's trajectories are not fixed.  There exist two scenarios depicted above. In the first scenario, $\dot{r}\geq0$, while in the second case $\dot{r}$ can be negative.}
  \label{Figure2}
\end{figure*}

\section{Geometric phase for communication between satellites}
\label{Section3}
Let us consider a communication link where Alice (sender) and Bob (receiver) are on different satellites located at radii of $r=a$ and $b$ from the centre of the Earth, respectively, where $a\leq b$. In this section, we assume that Alice and Bob are stationary with respect to the standard spherical coordinates of the Schwarzschild geometry. In next section, we utilise relativistic Doppler shift to generalise the result to the case that Alice and Bob are not stationary.  We use $\alpha$ to represent the angular separation of the two satellites; considering the lines from satellites to the centre of Earth, $\alpha$ is the angle between these two lines.\\
\noindent\textbf{First case, as shown in Fig.~\ref{Figure2}a for $\dot{r}\geq0$}:

The ``${+-+-}$'' component of the Riemann tensor evaluated on the geodesic is given in Eq.~\eqref{Rpmpm} which for large $r$ can be approximated to:
$R_{+-+-}= \frac{3 l^2}{2r^5}-\frac{1}{r^3}$. Applying the same approximation on $\dot{r}$ holds $\dot{r}= \sqrt{1 - \frac{l^2}{r^2}}$.
The geometrical factor defined in Eq.~\eqref{Eq32} then is given by
\begin{equation}
	\label{Eq36}
	{\cal G}=  \int_a^b \frac{dr}{\dot{r}} R_{+-+-}=\frac{\sqrt{b^2-l^2}}{2b^3}-\frac{\sqrt{a^2-l^2}}{2a^3}.
\end{equation}
Here $l$ is the minimum of the distance between the centre of the Earth and the line or the extrapolation of the line connecting Alice and Bob, i.e., $l = \frac{a b \sin \alpha}{\sqrt{a^2 + b^2 - 2 a b \cos \alpha}}$. The geometrical phase, Eq.~\eqref{Eq34}, is then given by,
\begin{equation}
	\label{Eq38}
	\chi_{g}=  + \frac{\omega_0  c m_\oplus (b^2-a^2)}{4  (\sigma a b)^2}  \sqrt{\frac{(a-b \cos\alpha)^2}{a^2+b^2 - 2 a b \cos\alpha }} \left(\frac{\sigma x^-}{c}\right)^2,
\end{equation}
where the Schwarzschild radius of the Earth $m_\oplus = \frac{2 G M_\oplus}{c^2}= 8.87~ \text{millimeters}$, and $c$ is recovered. Equation \eqref{Eq38} for $\alpha=0$ coincides to the result reported in~\cite{Exirifard:2020yuu} for radial communication between the Earth and the International Space Station.\\

\noindent\textbf{Second case as shown in Fig.~\ref{Figure2}b:} 
For $\alpha\geq  \arccos \frac{a}{b}$
as depicted in fig. \ref{Figure2}b, $\dot{r}$ can be negative in some parts of the geodesic. For the geometrical factor defined in \eqref{Eq32}, therefore, we can write:
\begin{eqnarray}
	\label{Eq39}
	{\cal G}&=& \int d\tau  R_{+-+-}= -\int_a^l \frac{dr}{|\dot{r}|} R_{+-+-}+\int_l^b \frac{dr}{|\dot{r}|} R_{+-+-}\nonumber\\
	&=&2 \int_l^a \frac{dr}{|\dot{r}|} R_{+-+-}+\int_a^b \frac{dr}{|\dot{r}|} R_{+-+-}
\end{eqnarray}
that leads to
\begin{equation}
	\label{Eq41}
	\chi_{g}=  - \frac{\omega_0  c m_\oplus (b^2+a^2)}{4  (\sigma a b)^2}  \sqrt{\frac{(a-b \cos\alpha)^2}{a^2+b^2 - 2 a b \cos\alpha }} (\frac{\sigma x^-}{c})^2.
\end{equation}

\section{On measuring the geometric phase near Earth}
\label{Section4}
In the following, we would like to evaluate the geometrical phase for a set of parameters to see if the geometrical phase can be detected in communication between two satellites around the Earth. 
In so doing, we first would like to generalise the result of the previous section to the case that Alice and Bob are not stationary.

Alice at position of ${r}_a$ prepares a time-bin Gaussian pulse with the mean frequency of $\omega_A$ and line-width of $\sigma_A$. The pulse that Alice produces in Alice's rest frame is given by:  
\begin{eqnarray}
	A_{\text{Alice}}  &=& A_{\text{Alice}}^0 e^{-\frac{(\sigma_{A} x^-)^2}{2}+ i \omega_{A} x^-}.
\end{eqnarray}  
Alice moves with velocity of $\vec{v}_{A}$ with respect to the local Riemann coordinates at $r_a$, which is stationary with respect to the standard spherical coordinates in the Schwarzschild geometry. In the local Riemann coordinates at $r_a$, since the source of the pulse moves with velocity of $\vec{v}$, the pulse at the event of its generation is given by,
\begin{eqnarray}
	A_{\text{a}}=A_{a}^0 e^{-\frac{(\sigma_{a} x^-)^2}{2}+ i \omega_{a} x^-}.
\end{eqnarray}
where $\omega_a = \Delta_{\vec{v}_a} \omega_{A}$, $\sigma_a = \Delta_{\vec{v}_a} \sigma_{A}$, and $\Delta_{\vec{v}_a}$ stands for the relativistic Doppler shift. We should still transform this pulse to the Fermi coordinates. Noticing the factor of $f$ and $1/f$ in the right-hand side of Eq.~\eqref{5a} and \eqref{5b}, the pulse in the Fermi coordinates, at the event of its generation, can be derived from the pulse in the local Riemann coordinates by scaling the frequencies: 
\begin{eqnarray}
	A_1  &=& A^0 e^{-\frac{(\sigma x^-)^2}{2}+ i \omega_0 x^-}.
\end{eqnarray}
where 
\begin{eqnarray}
\label{Eq40}
    \omega_0 
    =\Delta_{\vec{v}_a} f(r_a) \omega_{A},~\quad
    \sigma 
    =\Delta_{\vec{v}_a} f(r_a) \sigma_{A}\,.
\end{eqnarray}
Here $f=\sqrt{1-1/r}$, see the text after Eq.~\eqref{5c}. As the pulse moves toward Bob's geodesic, it gains an extra geometric phase. At the time of its detection, the pulse in the Fermi coordinates is given by,
\begin{eqnarray}
	A_1  &=& A^0 e^{-\frac{(\sigma x^-)^2}{2}+ i \omega_0 x^-} \exp(-i \chi_g),
\end{eqnarray}
where $\chi_g$ is given in Eq.~\eqref{Eq34}. Bob is moving with velocity $\vec{v}_B$ with respect to the local Riemann coordinates at $\vec{r}_b$. The pulse that Bob observes can be obtained by transforming the beam from the Fermi coordinates to the local Riemann coordinates, and then to the Bob's rest frame. Bob in his rest frame observes
\begin{eqnarray}
    A_{\text{Bob}}  &=& A^0 e^{-\frac{(\sigma_{B} x^-)^2}{2}+ i \omega_{B} x^-} \exp(-i \chi_g),
\end{eqnarray}
where 
\begin{eqnarray}
\label{Eq44}
    \omega_B =\frac{f(r_a)}{f(r_b)} \Delta_{\vec{v}_a}\Delta_{\vec{v}_b}\omega_{A},\quad
    \sigma_B =\frac{f(r_a)}{f(r_b)} \Delta_{\vec{v}_a}\Delta_{\vec{v}_b} \sigma_{A}\,.
\end{eqnarray}
Here, $ \Delta(\vec{v}_a)\Delta(\vec{v}_b)$ accounts for relativistic Doppler shift, while ${f(r_a)}/{f(r_b)}$ describes the gravitational red-shift. Equations~\eqref{Eq40} and \eqref{Eq44} can be utilised to re-express the geometric phase by:
\begin{equation}
    \chi_g = -\frac{\omega_B f(r_b) {\cal G}}{2\Delta_{\vec{v}_b}\sigma_B^2} (\sigma_B x^-)^2,
\end{equation}
where \eqref{Eq40} and \eqref{Eq44} are used to express $\omega_0$ in Eq.~\eqref{Eq34} in term of $\omega_B$. We notice that Bob can interpret $\chi_g$ as a time-dependent phase modulated over the Gaussian time-bin wave packet with the mean frequency $\omega_B$ and line-width of $\sigma_B$~\cite{hansen:01}. So, Bob can measure it. For terrestrial satellites with velocities less than $10^4$ mph, $|\Delta_{\vec{v}}-1|\leq 10^{-5}$. For satellites around the Earth,  $|f(r_b)-1|\leq 10^{-9}$. So in measuring the geometric phase  with a precision larger than $0.001$ percent, the Doppler and gravitational effects can be neglected, and the geometric phase that Bob observes can be approximated to
\begin{equation}
    \chi_g = -\frac{\omega  {\cal G}}{2\sigma^2} (\sigma x^-)^2
\end{equation}
where $\omega$ and $\sigma$  respectively respectively represent the mean frequency and the line-width. 

It is worth noting that in our notation, a plane-wave in $x^+$-direction is expressed as $e^{i\omega x^-}= e^{i\frac{\omega}{\sqrt{2}} (x^3 - ct)}$. In optics, however, a plane-wave in $x^3$-direction is represented by $e^{i\omega (x^3- ct)}$. Thus, what we define as a frequency is $\sqrt{2}$ times the notation in optics. The geometrical phase presented in Eq.~\eqref{Eq38} in the standard optic notation is $\chi_{g1}^s=  f_1(\alpha)\, y^2 $, where 
\begin{subequations}
\begin{eqnarray}
	\label{Eq7.10}
	y &=& \frac{\sigma (x^3- ct)}{c},\\
	\label{Eq7.11}
	f_1(\alpha) &=&\! \frac{\sqrt{2}\nu_0  c m_\oplus (b^2-a^2)}{16 \pi  (\sigma a b)^2} \! \sqrt{\frac{(a-b \cos\alpha)^2}{a^2+b^2\! - 2 a b \cos\!\alpha }}, \quad
\end{eqnarray}
\end{subequations}
where $\omega_0$ is replaced with ${\sqrt{2}\nu^s_0}/{2 \pi}$, instead of ${\nu_0}/{2 \pi}$, and $\sigma$ is changed to $\sqrt{2}\sigma$. The geometrical factor presented in Eq.~\eqref{Eq41} in the standard optical notation reads $\chi_{g2}^s= f_2(\alpha)\,y^2$, where
\begin{equation}
	\label{Eq7.13}
	f_2(\alpha) = -  \frac{\sqrt{2}\nu_0  c m_\oplus (b^2+a^2)}{16 \pi  (\sigma a b)^2}  \sqrt{\frac{(a-b \cos\alpha)^2}{a^2+b^2 - 2 a b \cos\alpha }},
\end{equation}
and $y$ is defined in Eq.~\eqref{Eq7.10}. In order to understand how the geometric factor depends on the Newton Gravitational constant, Heisenberg constant and speed of light in vacuum, we treat the photon as a particle entity with an energy of $E_0=\hbar \nu_0$ and variance of $\Delta E_0=\hbar\sigma$. The geometric phase, then, can be expressed by,
\begin{equation}
    \chi_g = l_p^2  \tilde{f}(a, b,\alpha) \times\frac{(M_\oplus c^2) E_0}{(\Delta E_0)^2},
\end{equation}
where $l_p^2= \frac{G_N \hbar}{c^3}$ is the Planck's length, while $\tilde{f}(a, b,\alpha)$ encodes the geodesic's details and has the unit dimension of the inverse length squared, and $\frac{(M_\oplus c^2) E_0}{(\Delta E_0)^2}$ encodes properties of the pulse. Near the Earth, $l_p^2 \tilde{f}$ can be estimated to be at \href{https://www.wolframalpha.com/input/?i=(Planck length)^2/(10000 km)^2}{the order of $10^{-84}$}. The factor of $\frac{(M_\oplus c^2) E_0}{(\Delta E_0)^2}$ can become arbitrary large in the limit of $\Delta E_0 \to 0$, but this divergence points to the break of the perturbative methods in calculating the geometric phase and demands non-perturbative derivation of the geometric phase.  The ultra-stable lasers with bandwidth of $5$ mHz  at $194$ THz reported in \cite{Zhang:2017dya,Matei:2017kug, 2012NaPho...6..687K} gives rise to  $\frac{(M_\oplus c^2) E_0}{(\Delta E_0)^2}$ at \href{https://www.wolframalpha.com/input/?i=(Mass+of+Earth*c^2)(hbar*194+THz)/(hbar*0.005+Hz)^2}{the order of $10^{94}$}. We, however, notice that $\sigma$ should satisfy some other conditions: Here, $\delta\chi$ is calculated by perturbative methods, so a value should be chosen for $\sigma$ that results in $|\chi_g|\leq 1$. The length of the wave packet in the direction of the propagation is given by $\frac{c}{\sigma}$, and the employed method has assumed that the Riemann tensor is constant within the wave packet. We also have implicitly assumed that the whole of the wave packet propagates in the space before its detection. The choice of bandwidth of $5$ mHz at $194$ THz violates these conditions. However, bandwidth at a few kHz satisfies these conditions and (as shown below) leads to a measurable value of $\chi_g$.
\begin{figure}[t]
	%
	\begin{center}
		\includegraphics[width=0.45 \textwidth]{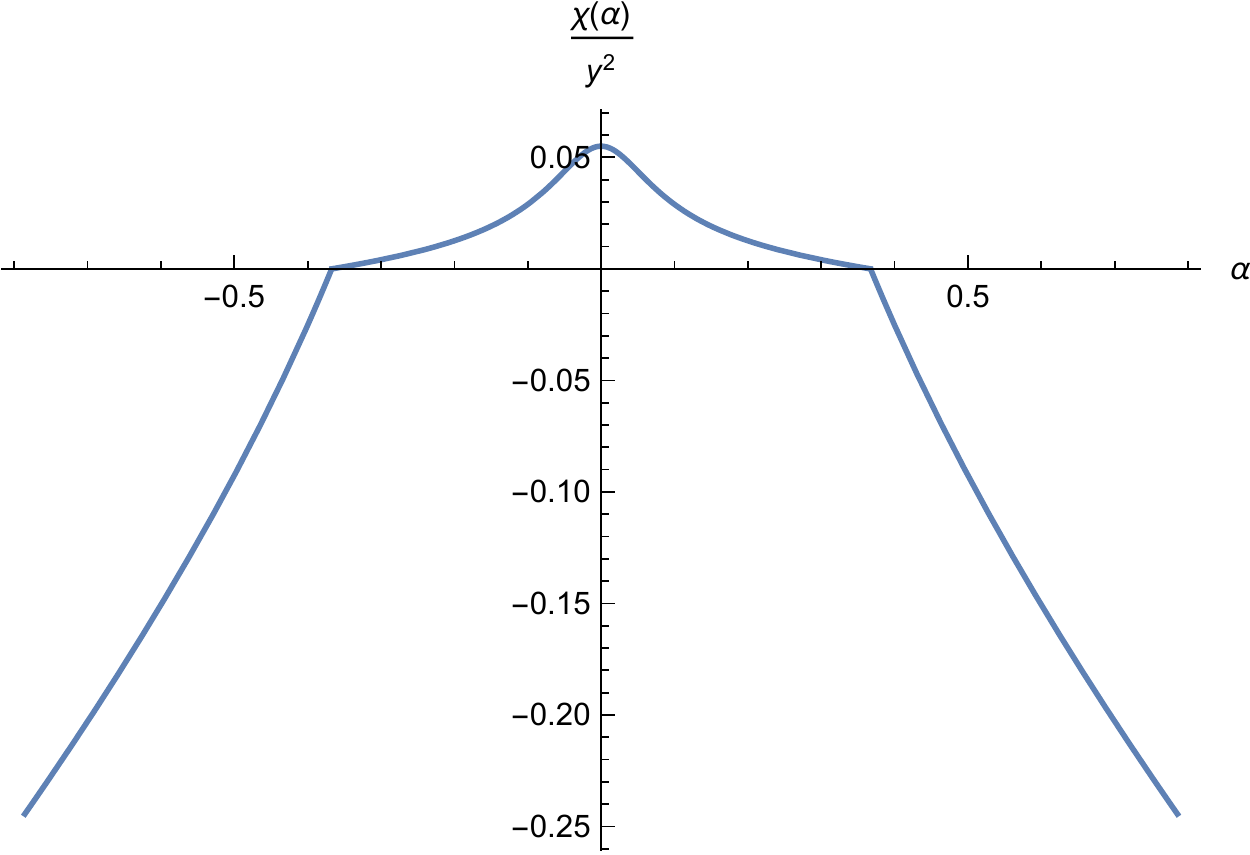}
	\end{center}
\caption{
The geometrical phase is a time-dependent phase quadratic in  $y=\frac{\sigma (x^3- ct)}{c}$   modulated on a time-bin Gaussian wave-packet. The coefficient of the quadratic term depends on the distance between two satellites and their apparent angle as seen from the centre of the Earth. This figure depicts the dependency of the coefficient of the quadratic term  in the geometrical factor for satellites at $a=7,000$~km and $b=7,500$~km, and for Gaussian time-bin communication performed with $\nu_0=2.87 \times 10^{15}$~Hz, $\sigma= 3.16$~kHz . The vertical axis is the coefficient of the quadratic term in the geometric factor $\frac{\chi_g}{y^2}$; the horizontal axis is the apparent angle.}
\label{fig3}	
\end{figure}

In order to consistently neglect the effect of atmosphere on the geometrical factor, let us consider communication between satellites in space, and choose $a= 7,000$~km and $b= 7,500$~km, respectively. We notice that  commercial portable continuous lasers with a line-width of $1$~Hz  at the wavelength of $657$ ~nm exists~\footnote{\href{https://www.menlosystems.com/products/ultrastable-lasers/}{https://menlosystems.com/products/ultrastable-lasers/}}. These can be used to construct a sharp Gaussian time-bin with line-width of  $\sigma=3.16$ kHz for $\nu_0=2.87 \times 10^{15}$~Hz. Using these numerical values simplify $f_1$ and $f_2$ to:
\begin{subequations}
\begin{eqnarray}
	\label{Eq7.18}
	f_1(\alpha) &=&  \frac{0.055\times |7.5 \cos\alpha-7|}{\sqrt{105.25- 105 \cos \alpha}},\\
	\label{Eq7.19}
	f_2(\alpha) &=&-  \frac{0.802 \times |7.5 \cos\alpha-7|}{\sqrt{105.25- 105 \cos \alpha}}. 
\end{eqnarray}
\end{subequations}
We do not want the signal to enter the atmosphere below the altitude of $400$~km, where the air density is about $10^{-12}\,{\text{kg}}/{\text{m}^3}$. Thus, the gravitational effects are a couple of orders larger than the atmosphere's diffraction. This constrains $\alpha$ to
$|\alpha| \leq 
0.675$
, where the saturation occurs when the line connecting the satellites is tangent to the orbit with a radius of $6800$ km. Figure~\ref{fig3} depicts the value of the geometrical phase divided by $y^2$ for the allowed range of $\alpha$. The dependency on $\alpha$ is very non-trivial. For the chosen parameters, the value of ${\chi[\alpha]}/{y^2}$ ranges from $0.05$ to $-0.187$. This means that the geometrical phase at $y=1$, depending on the value of $\alpha$, varies in the range of $0.05$ to $-0.187$ Radians which can be measured. One may choose smaller values of $\sigma$ to obtain a larger geometrical phase. It, however, should be noted that the geometrical phase is calculated by perturbative methods. So the choices leading to a large geometric factor can not be supported by perturbative methods developed in this work. The choice of $\sigma=3.16$ kHz for $\nu_0=2.87 \times 10^{15}$~Hz, and $a=7,000$ km and $b=7,500$ km as depicted in Figure \ref{fig3} leads to a measurable geometric phase consistent with a perturbative calculation.

\section{Conclusions}
\label{Section5}
As the photon's wave-function travels along a null geodesic, it interacts with the Riemann tensor around the geodesic. The interaction distorts the photon's wave-function. Here, the Fermi coordinates along the null geodesic have been utilized. The equations for the $U(1)$ gauge field theory in the Fermi coordinates have been calculated. The equation for the interaction between the Riemann tensor and the photon's wave-function has been derived and mapped to a time-dependent Schr\"odinger equation in $(2+1)$ dimensions.  It has been shown that as a Gaussian time-bin wave-packet, with a sharp width of $\sigma$ around the frequency of $\omega_0$ travels over the null geodesic, it gains an extra geometric phase given by 
$\chi_{g}=-\frac{\omega_0 {\cal G}}{2 \sigma^2} (\sigma x^-)^2$, where $\cal G$ is  ``${+-+-}$'' component of the Riemann tensor in the Fermi coordinates evaluated on and integrated over the null geodesic:
${\cal G}= \int R_{+-+-}(x^+) dx^+$, where $x^+$ represents the coordinate of Fermi frame tangent to the central null geodesic, and  the integration is performed from the event of generation of the pulse to the event of its detection.

The space-time geometry outside the Earth has been approximated by the Schwarzschild space-time geometry.  The geometrical phase has been calculated for a signal sent between two satellites, located at radii of $a$ and $b$, respectively. The current commercial ultra-stable continuous-wave lasers (wavelength of $657$ nm and $\sigma=3.16$ kHz) have been utilized to calculate the geometrical phase   between satellites at radii $7,000$~km and $7,500$~km. It has been shown that for the chosen range of the parameters, the geometrical phase within the peak of the Gaussian pulse varies from $0.05$ to $-0.187$ Radians, as depicted in Fig.~\ref{fig3}. This illustrates that the predicted geometrical phase can be measured by the currently available commercial devices. The geometrical phase calculated in the current work is consequent of applying quantum field theory in curved space-time geometry. The three paradigms of special relativity, general relativity, and quantum mechanics are equally important in this derivation. It, therefore, is a prediction of how Einstein's gravity ``talks" to the quantum realm. Hence, measuring this phase will provide the first experimental datum on \textit{if} and \textit{how} gravity affects the quantum realm. 

\acknowledgments
This work was supported by the High Throughput and Secure Networks Challenge Program at the National Research Council of Canada, the Canada Research Chairs (CRC) and Canada First Research Excellence Fund (CFREF) Program, and Joint Centre for Extreme Photonics (JCEP). We thank 
Felix Hufnagel for proofreading the paper.

\end{document}